# Proton Transport through One Atom Thick Crystals

S. Hu[1,2], M. Lozada[1], F. C. Wang[3], A. Mischenko[1], F. Schedin[2], R. R. Nair[1], E. W. Hill[2], D. V. Boukhvalov[4], M. I. Katsnelson[4], R. A. W. Dryfe[5], I. V. Grigorieva[1], H. A. Wu[3], A. K. Geim[1,2]

[1]School of Physics & Astronomy, University of Manchester, Manchester, M13 9PL, UK
[2]Manchester Centre for Mesoscience & Nanotechnology, Manchester M13 9PL, UK
[3]Chinese Academy of Sciences Key Laboratory of Mechanical Behavior and Design of Materials, Department of Modern Mechanics, University of Science and Technology of China, Hefei, Anhui 230027, China
[4]Institute for Molecules and Materials, Radboud University of Nijmegen, 6525 AJ Nijmegen, The Netherlands
[5]School of Chemistry, University of Manchester, Manchester, M13 9PL, UK

**Graphene is impermeable to all gases and liquids[1-3], and even such a small atom as hydrogen is not expected to penetrate through graphene's dense electronic cloud within billions of years[3-6]. Here we show that monolayers of graphene and hexagonal boron nitride (hBN) are unexpectedly permeable to thermal protons, hydrogen ions under ambient conditions. As a reference, no proton transport could be detected for a monolayer of molybdenum disulfide, bilayer graphene or multilayer hBN. At room temperature, monolayer hBN exhibits the highest proton conductivity with a low activation energy of ≈0.3 eV but graphene becomes a better conductor at elevated temperatures such that its resistivity to proton flow is estimated to fall below $10^{-3}$ Ohm per $cm^2$ above 250°C. The proton barriers can be further reduced by decorating monolayers with catalytic nanoparticles. These atomically thin proton conductors could be of interest for many hydrogen-based technologies.**

Graphene has recently attracted renewed attention as an ultimately thin membrane that can be used for development of novel separation technologies (for review, see refs. 7,8). If perforated with atomic or nanometer accuracy, graphene may provide ultrafast and highly selective sieving of gases, liquids, ions, etc.[2,9-19] However, in its pristine state, graphene is absolutely impermeable for all atoms and molecules moving at thermal energies[1-7]. Theoretical estimates for the kinetic energy $E$ required for an atom to penetrate through monolayer graphene vary significantly, depending on the employed model, but even the smallest literature value of 2.4 eV for atomic hydrogen[3-6] is 100 times larger than typical $k_BT$ which ensures essentially an impenetrable barrier ($k_B$ is the Boltzmann constant and $T$ the temperature). Therefore, only accelerated atoms are capable of penetrating through the one atom thick crystal[20,21]. The same is likely to be valid for other two dimensional (2D) crystals[22,23], although only graphene has so far been considered in this context. Protons can be considered as an intermediate case between electrons that tunnel relatively easily through atomically thin barriers[24] and small atoms. It has been calculated that $E$ decreases by a factor of up to 2 if hydrogen is stripped of its electron[4,5]. Unfortunately, even the latter barrier is still prohibitively high to allow appreciable transport of thermal protons ($E \approx 1.2$ eV is estimated[5] to result in permeation rates of $\sim 10^9$ sec).

Despite the pessimistic prognosis, we have investigated the possibility of proton permeation through monocrystalline membranes made from mono- and few- layers of graphene, hBN and molybdenum disulfide ($MoS_2$). The 2D crystals were obtained by mechanical cleavage and then suspended over micrometer size holes etched through $Si/SiN_x$ wafers. Details of fabrication procedures are described in Supplementary Information, section 1. The resulting free-standing membranes were checked for the absence of pinholes and defects (Supplementary Information, sections 1-3) and spin coated from both sides with Nafion, a polymer that exhibits high proton and negligible electron conductivity[25]. Finally, two proton injecting $PdH_x$ electrodes[26,27] were deposited onto Nafion from both sides of the wafer (Supplementary Figs 1 and 2). As illustrated in the left inset of Fig. 1a, 2D crystals effectively serve as atomically thin barriers between two Nafion spaces. For electrical measurements, samples were placed in a hydrogen-argon atmosphere at 100% humidity, which ensured high conductivity of Nafion films[25,26]. Examples of I-V characteristics measured for devices incorporating monolayers of graphene, hBN and $MoS_2$ are shown in Fig. 1a. This behavior is highly reproducible, as illustrated by statistics in Fig. 1b for a number of different membranes. The measured proton current $I$ is found to vary linearly with bias $V$, and the conductance $S = I/V$ to be proportional to the membrane area $A$ (Supplementary Figs 3-5). For devices prepared in the same manner but without a 2D membrane ('bare hole'), $S$ was ~50 times higher than in the presence of monolayer hBN (Supplementary Fig. 3). This ensures that the measured areal conductivity $\sigma = S/A$ is dominated by the 2D crystals and that Nafion gives rise only to a relatively small series resistance. In the opposite limit of thick barriers (e.g., a few nm thick graphite or thick metal or dielectric films evaporated between the Nafion spaces), we find a parasitic parallel conductance of ~10 pS, which could be traced back to leakage currents along $SiN_x$ surfaces in high humidity. Within this accuracy, we could not detect any proton current through monolayer $MoS_2$, bilayer graphene, tetra-layer hBN or thicker 2D crystals.

The difference in permeation through different 2D crystals can qualitatively be understood if we consider the electron clouds that have to be overcome by passing protons. One can see from the insets of Fig. 1b that monolayer hBN is more 'porous' than graphene, reflecting the fact that the boron nitride bond is strongly polarized with valence electrons concentrated around nitrogen atoms. For $MoS_2$, the cloud is much denser because of the larger atoms involved (Supplementary Fig. 8). The absence of detectable $\sigma$ for bilayer graphene can be attributed to its AB stacking such that 'pores' in the electron cloud in one layer are covered by density maxima within the other layer. In contrast, hBN crystals exhibit the AA' stacking, which leads to an increase in the integrated electron density with increasing number of layers but allows the central pore in the cloud to persist even for multilayer hBN membranes.

It is instructive to emphasize that there is no correlation between proton and electron transport through 2D crystals. Indeed, hBN exhibits the highest proton conductivity but is a wide gap insulator with the highest tunnel barrier[23,24]. In contrast, monolayer $MoS_2$ that shows no discernable proton permeation is a heavily doped semiconductor with electron-type conductivity[22,28]. Furthermore, numerous studies using transmission and tunneling microscopy and other techniques have so far failed to find even individual pinholes (atomic-scale defects) in graphene and hBN prepared using the same cleavage technique as employed in the present work (see, e.g., refs 1,2,24). Similar examination of our specific membranes is described in Supplementary Information, section 3. In contrast, MoS$_2$ monolayers contain a high density of sulfur vacancies[29] but nonetheless exhibit little proton conductivity. These observations combined with the high reproducibility of our measurements for different devices, the linear scaling with $A$ and the consistent behavior with increasing the number of layers assure that the reported σ represent the intrinsic proton properties of the studied membranes.

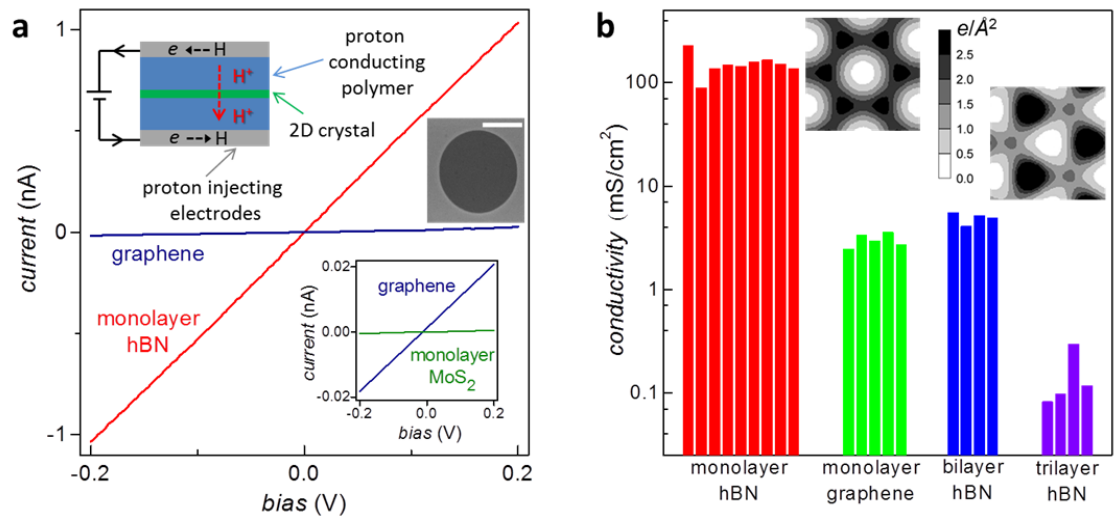

**Figure 1 | Proton transport through 2D crystals. a**, Examples of I-V characteristics for monolayers of hBN, graphite and MoS$_2$. The upper inset shows experimental schematics. Middle inset: Electron micrograph of a typical graphene membrane before depositing Nafion. Scale bar: 1 µm. In a scanning electron microscope, 2D crystals give rise to a homogenous dark background and can only be seen if contamination, defects or cracks are present (Supplementary Fig. 2). Small (pA) currents observed for MoS$_2$ membrane devices (lower inset) are due to parasitic parallel conductance. **b**, Histograms for 2D crystals exhibiting detectable proton conductivity. Each bar represents a different sample with a 2 µm diameter membrane. Left and right insets: charge density (in electrons per Å$^2$) integrated along the direction perpendicular to graphene and monolayer hBN, respectively. The white areas are minima at the hexagon centers; the maxima correspond to positions of C, B and N atoms.

To determine the barrier heights $E$ presented by graphene and hBN, we have measured $T$ dependences of their σ (Fig. 2a) which are found to exhibit the Arrhenius-type behavior, $\exp(-E/k_BT)$. Note that conductivity of Nafion not only contributes little to the overall value of $S$ but also changes only by a factor of ~1.5 for the same $T$ range (Supplementary Fig. 5). The activation behavior yields $E$ = 0.78±0.03, 0.61±0.04 and 0.3±0.02 eV for graphene, bilayer hBN and monolayer hBN, respectively. The proton barrier for graphene is notably lower than the values of 1.2–2.2 eV, which were found using ab initio molecular dynamics simulations and the climbing image nudged elastic band method[4-6]. We have reproduced those calculations for graphene and extended them onto monolayer hBN (Supplementary Information, section 5). Our results yield $E$ =1.25–1.40 for graphene, in agreement with refs 4-5, and ≈0.7 eV for monolayer hBN. The disagreement between the experiment and theory in the absolute value of $E$ is perhaps not surprising given the complex nature of possible pathways and sensitivity of the calculations to pseudopotentials, the exchange-correlation functional, etc. Alternatively, the difference can arise due to the fact that protons in Nafion/water (Supplementary Information, section 6) move along hydrogen bonds[25] rather than in vacuum as the theory has assumed so far.

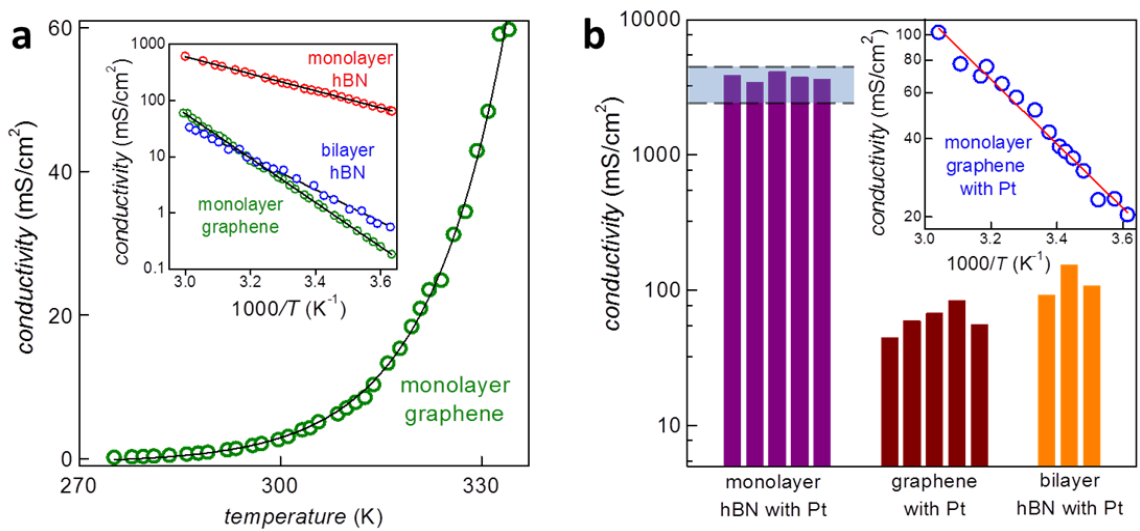

*Figure 2 | Proton barrier heights and their catalytic suppression. **a**, T dependences of proton conductivity for 2D crystals. The inset shows log($\sigma$) as a function of inverse T. Symbols are experimental data; solid curves are the best fits to the activation dependence. The T range is limited by freezing of water in Nafion, and we normally avoided T above 60°C to prevent accidental damage because of different thermal expansion coefficients. **b**, Proton conductivity is strongly enhanced if 2D crystals are decorated with catalytic nanoparticles. Each bar is a different device. The shaded area indicates the conductivity range found for bare-hole devices (Nafion/Pt/Nafion: no 2D crystal but for full similarity the same amount of Pt was evaporated). Inset: Arrhenius-type behavior for graphene with Pt, yielding E =0.24±0.03 eV. Monolayer hBN with Pt exhibits only a weak T dependence (Supplementary Fig. 5) which indicates that the barrier becomes comparable to $k_BT$.*

For certain applications, it is desirable to achieve the highest possible proton conductivity. For example, hydrogen fuel cells require membranes with $\sigma$ >1 S per cm$^2$. This condition is satisfied by monolayers of hBN and graphene above 80 and 110 °C, respectively (inset of Fig. 2a). Moreover, graphene remains stable in oxygen and humid atmosphere up to 400°C (ref. 30), and the extrapolation of our results to 'very safe' 250°C yields extremely high $\sigma$ >10$^3$ S/cm$^2$. Furthermore, noticing that platinum group metals have a high affinity to hydrogen, we have investigated their influence on proton transport through 2D crystals. To this end, a discontinuous layer of Pt or Pd (nominally, 1-2 nm thick) was evaporated onto one of the surfaces of 2D crystals (Supplementary Information). Fig. 2b shows that the added catalytic layer leads to a significant increase in $\sigma$. For monolayer hBN, the measured S becomes indistinguishable from that of reference 'bare hole' devices (Fig. 2b). This shows that our measurements become limited by Nafion's series resistance and Pt-activated monolayer hBN is no longer a bottleneck for proton permeation. On the other hand, for graphene and bilayer hBN activated with Pt, the series resistance remains relatively small and the measurements still reflect their intrinsic properties. By studying $\sigma(T)$, we find that Pt reduces the activation energy E by as much as ≈0.5 eV to ≈0.24 eV (Fig. 2b). Our simulations of the catalytic effect yield a reduction in E by ≈0.65 eV, in qualitative agreement with the experiment (Supplementary Information). The mechanism behind this barrier reduction can be attributed to attraction of passing protons to Pt (Supplementary Fig. 10). Note that the measurements in Fig. 2b set only a lower limit of ≈3 S/cm$^2$ on room-$T$ conductivity of catalytically-activated monolayer hBN and, if the membranes experience qualitatively similar reduction in $E$ as observed for graphene, we expect essentially a barrier-less proton transport. It would require membranes with much larger area to determine intrinsic σ for catalytically-activated hBN.

Finally, we demonstrate directly that the observed electric currents are due to proton flux through the 2D membranes. To this end, we have prepared devices such as shown in the insets of Fig. 3. Here, one of the Nafion/PdH$_x$ electrodes is removed, and the graphene surface decorated with Pt faces a vacuum chamber equipped with a mass spectrometer. If no bias is applied between graphene and the remaining PdH$_x$ electrode, we cannot detect any gas leak (including He) between the hydrogen and vacuum chambers. Similarly, no gas flow could be detected for positive bias on graphene. However, by applying a negative bias we have measured a steady H$_2$ flux into the vacuum chamber. Its value is determined by the number of protons, $I/e$, passing through the membrane per sec. Using the ideal gas law, one can easily derive the relation $F = k_BT(I/2e)$ where the flow rate $F$ is the value measured by the mass spectrometer tuned to molecular hydrogen. The latter dependence is shown in Fig. 3 by the solid red line, in excellent agreement with the experiment.

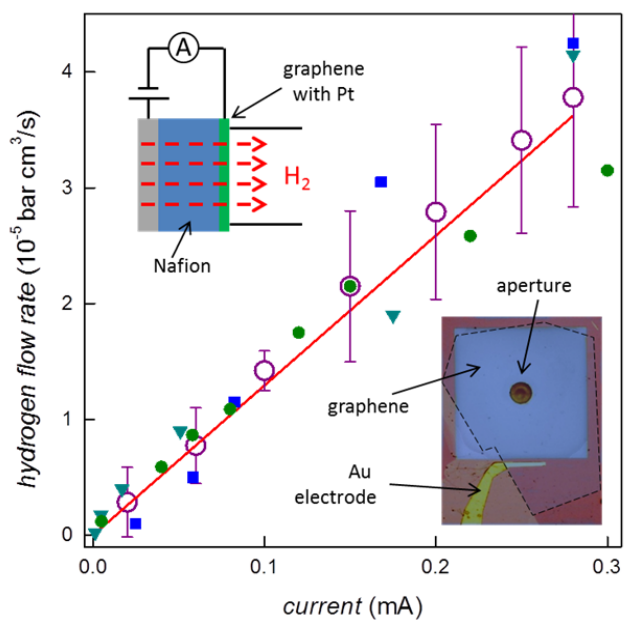

*Figure 3 | Current-controlled proton flux. Top inset: Schematics of our experiment. Monolayer graphene decorated with Pt nanoparticles separates a vacuum chamber from the Nafion/PdH$_x$ electrode placed under the same H$_2$/H$_2$O conditions as described above. Protons permeate through the membrane and rapidly recombine on the other side into molecular hydrogen. The hydrogen flux is detected by a mass spectrometer (Inficon UL200). Different symbols refer to different devices, error bars are shown for one of them, and the red line is the theoretically expected flow rate. Bottom inset: Optical image of one of the devices. Graphene contoured by the dashed lines seals a circular aperture of 50 μm in diameter. Nafion is underneath the graphene membrane.*

In conclusion, monolayers of graphene and hBN represent a new class of proton conductors. In addition to the fundamental interest in how subatomic particles transfer through atomically thin electron clouds, the membranes can find use in various hydrogen technologies. For example, 2D crystals can be considered as proton membranes for fuel cells. They are highly conductive to protons and chemically and thermally stable and, at the same time, impermeable to $H_2$, water or methanol. This could be exploited to solve the problem of fuel crossover and poisoning in existing fuel cells. The demonstrated current-controlled source of hydrogen is also appealing at least for its simplicity and, as large-area graphene and hBN films are becoming commercially available, the scheme may be used to harvest hydrogen from gas mixtures or air.

## Supplementary Information

**#1 Experimental devices**

Figure S1 explains our microfabrication procedures. We start with preparing free-standing silicon nitride ($SiN_x$) membranes from commercially available Si wafers coated from both sides with 500 nm of $SiN_x$. Reactive ion etching (RIE) is employed to remove a 1×1 mm² section from one of the $SiN_x$ layers (steps 1&2 in Fig. S1). The wafer is then exposed to a KOH solution that etches away Si and leaves a free-standing $SiN_x$ membrane of typically 300×300 µm² in size (step 3). During step 4, a circular hole is drilled by RIE through the $SiN_x$ membrane using the same procedures as in steps 1&2. Next, a 2D crystal (graphene, hBN or $MoS_2$) is prepared by the standard micromechanical exfoliation [S1] and transferred on top of the membrane using either the wet or dry technique [S2,S3] to cover the aperture in $SiN_x$ (step 5).

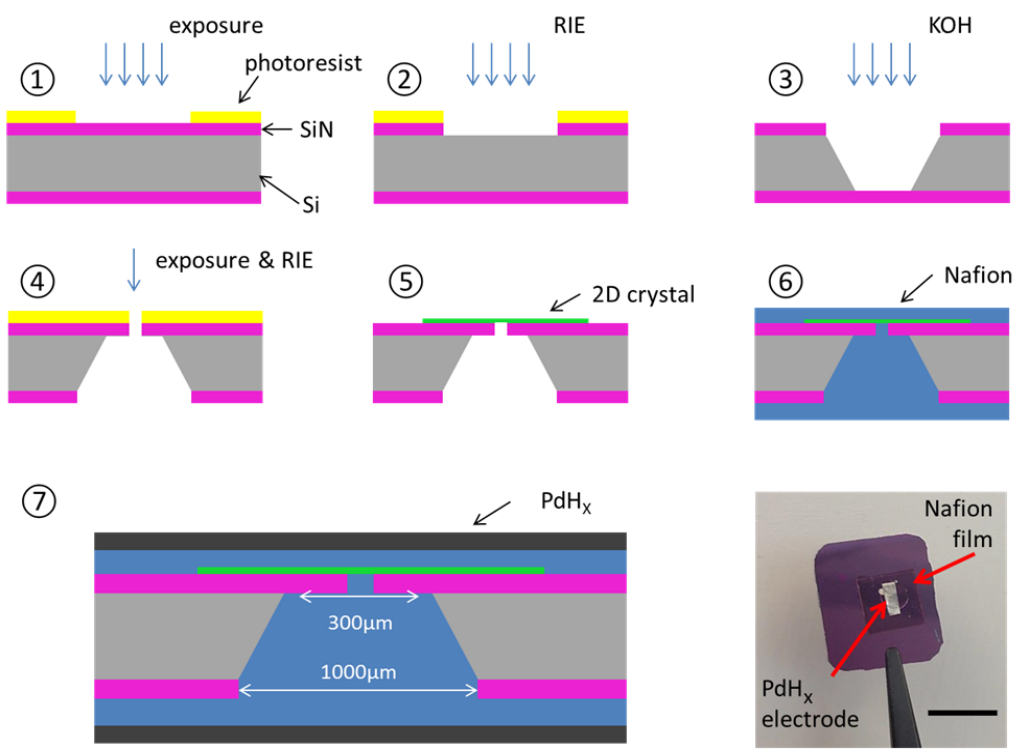

**Figure S1 | Microfabrication process flow.** ① An etch mask is made by photolithography. ② RIE is used to remove the exposed $SiN_x$ layer. ③ Si underneath is etched away by wet chemistry. ④ By repeating steps 1 and 2, a hole is drilled through the membrane. ⑤ 2D crystal is transferred to cover the hole. ⑥ Nafion is deposited on both sides of the wafer. ⑦ Palladium-hydride electrodes are attached. Bottom right: Optical photo of the final device. Scale bar: 1 cm.

After step 5, the suspended membranes could be examined for their integrity and quality in a scanning electron microscope (SEM). As mentioned in the main text, pristine 2D crystals give little SEM contrast, and it requires some contamination to notice 2D membranes on top of the holes. Contamination can be accidental as in the case of Fig. S2a or induced by the electron beam (Fig. S2b). If cracks or tears are present, they are clearly seen as darker areas (inset of Fig. S2b).

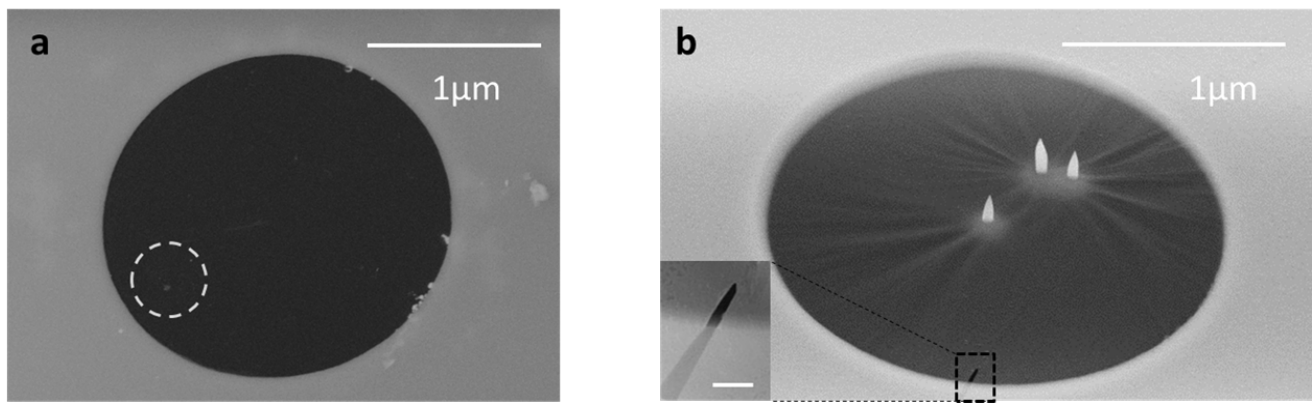

*Figure S2 | SEM images of suspended 2D membranes. **a**, Monolayer graphene with some accidental contamination. One of the particles away from the edge is marked by the white circle. **b**, Suspended graphene with pillars of hydrocarbon contamination intentionally induced by a focused electron beam. The inset shows a crack in the membrane; scale bar: 100 nm.*

The fabrication of devices for electrical measurements continues with depositing a proton-conducting polymer layer. A Nafion 117 solution (5%) is drop-cast or spin-coated on both sides of a suspended 2D membrane (step 6 in Fig. S1). Finally, palladium hydride (PdH$_x$) electrodes are mechanically attached to the Nafion layers. To synthesize such electrodes, a 25 μm thick Pd foil is left overnight in a saturated hydrogen-donating solution following the recipe reported in ref. S4. This leads to atomic hydrogen being absorbed into the crystal lattice of Pd turning it into PdH$_x$. The resulting devices are placed in a water saturated environment at 130°C to crosslink the polymer and improve electrical contacts.

The described experimental design is optimized to take into account the following considerations. First, electric currents in Nafion are known to be carried out exclusively by protons that hop between immobile sulfonate groups [S5]. Nafion is not conductive for electrons, which can be evidenced directly by, for example, inserting a gold film across a Nafion conductor, which then breaks down the electrical connectivity. Accordingly, protons are the only mobile species that can pass between PdH$_x$ electrodes. Second, PdH$_x$ is widely used as a proton injecting material that converts an electron flow into a proton one by the following process: $PdH_x \rightarrow Pd + xH^+ + xe^-$ [S6-S8]. This property combined with a large area of our electrodes (relatively to the membrane area $A$) makes the contact resistance between Nafion and PdH$_x$ negligible so that the circuit conductance in our experiments is limited by either 2D crystals or, in their absence, by the Nafion constriction of diameter $D$.

For the catalytically-activated measurements, 1-2 nm of Pt were deposited by e-beam evaporation directly onto the suspended membrane to form a discontinuous film prior to the Nafion coating. Thicker, continuous films were found to block proton currents, which could be witnessed as numerous hydrogen bubbles that appeared under Pt after passing electric current. Typically, our Pt films resulted in ~80% area coverage, which reduced the effective area for proton transport accordingly, as found by depositing such films between Nafion spaces but without 2D membranes (see below). Pd films were found to be less blocking and continuous films up to 10 nm in thickness did not significantly impede the proton flow. Otherwise, both Pd and Pt films resulted in similar enhancement of proton transport through 2D crystals.

#2 Electrical measurements

The devices described above were placed inside a chamber filled with a forming gas (10% $H_2$ in argon) and containing some liquid water to provide 100% relative humidity. I-V curves were recorded by using DC measurements. We varied voltage in a range of typically up to 1 V at sweep rates up to 0.5 V/min. Under these conditions, the curves were non-hysteretic and highly reproducible. The devices were stable for many weeks if not allowed to dry out.

To characterize our experimental setup, we first measured leakage currents in the absence of a proton conductive path. To this end, two metallic contacts were placed onto the opposite surfaces of a piece of a fresh Si/SiN$_x$ wafer and I-V characteristics were measured under the same humid conditions. Conductance of the order of ~5 pS was normally registered. We also used fully processed devices and then mechanically removed the Nafion film and electrodes. In the latter case, the parasitic conductance was slightly (by a factor of 2) higher, which is probably due to a residue left of SiN$_x$ surfaces during processing. In principle, it would be possible to reduce the leakage currents by using, for example, separate chambers at the opposite sides of the Si wafer [S9] but the observed parasitic conductance was deemed small enough for the purpose of the present work.

As a reference, we studied conductivity of 'bare-hole' devices that were prepared in exactly the same manner as our membrane devices but without depositing a 2D crystal to cover the aperture (step 5 in Fig. S1 is omitted). Figure S3 shows conductance of such devices as a function of their diameter $D$. Within the experimental scatter, conductance $S$ increases linearly with $D$, in agreement with Maxwell's formula: $S = \sigma_N D$ [S10]. The latter is derived by solving Laplace's equation for two semi-spaces that have conductivity $\sigma_N$ and are connected by a hole with $D$ much larger than the length $d$ of the opening. In our case, $d$ =500 nm and the condition is comfortably satisfied, except for possibly the smallest membranes in Fig. S3 with $D$ =2 μm.

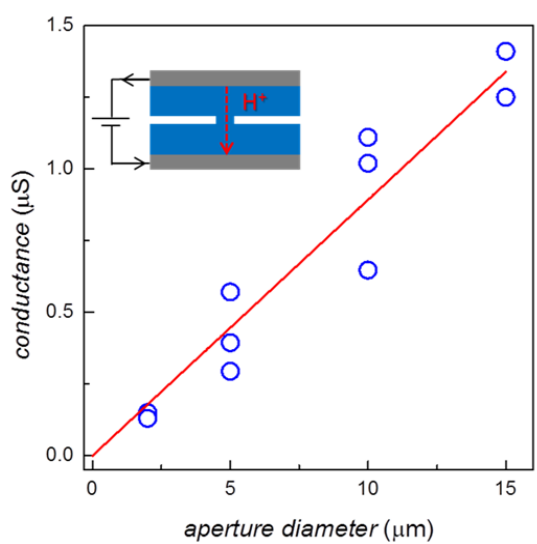

*Figure S3 | Bare-hole devices with different apertures.* *Their conductance exhibits a linear dependence on D as expected for this geometry. The inset illustrates schematics of such devices.*

From the dependence shown in Figure S3, we can estimate conductivity of our Nafion films as ≈1 mS/cm. As discussed in the main text, Nafion's conductivity did not limit our measurements of proton transport through 2D crystals, except for the case of catalytically-activated monolayer hBN. Nonetheless, we note that the found $\sigma_N$ is two orders of magnitude smaller than values achievable for highest-quality Nafion [S11]. There are two reasons for this. First, solution-cast Nafion is known to lose typically one order of magnitude in conductivity [S12,S13]. Second, Nafion is normally pretreated by boiling in $H_2O_2$ and $H_2SO_4$ for several hours [S11-S13]. If the latter procedure was used, our Nafion films indeed increased their conductivity by a factor of 10, reaching the standard values for solution-cast Nafion of ~10 mS/cm. Unfortunately, this harsh treatment could not be applied to our membrane devices that became destroyed with Nafion films delaminating from $SiN_x$.

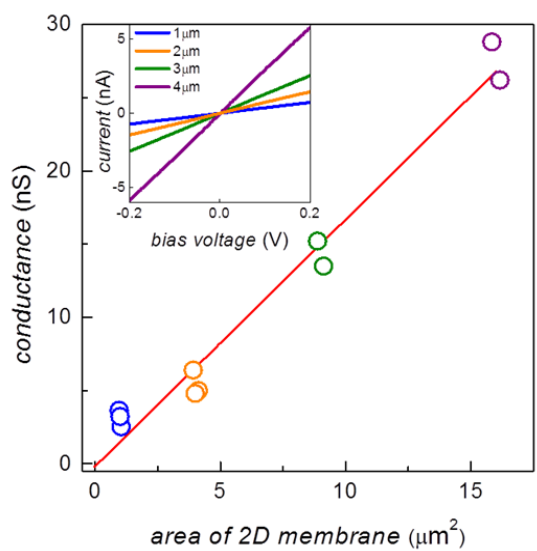

*Figure S4 | Proton conductance through monolayer hBN membranes of different sizes.* *Conductance scales quadratically with D, that is, linearly with A. Inset shows I-V characteristics for devices with different D.*

For consistency, most of the 2D membranes reported in the main text were made 2 μm in diameter. However, we also studied many other membranes with diameters ranging from 1 to 50 μm. We found that their conductance scaled linearly with the aperture area *A*. Figure S4 shows this for 10 monolayer hBN devices with *D* between 1 and 4 μm. Within the typical experimental scatter for devices with the same *D*, the conductance increases linearly with the area *A* of 2D membranes, in agreement with general expectations. The same scaling was also observed for graphene membranes.

Finally, we have reported in the main text that proton conductivity of catalytically-activated monolayer hBN is so high that the series resistance of Nafion becomes the limiting factor in our measurements. This is further evidenced by comparing *T* dependences of different devices in which Nafion was the limiting factor. Those include 'bare-hole' devices (Nafion only), 'bare-hole' devices with Pt (Nafion/Pt/Nafion) and monolayer hBN membranes activated with Pt. Figure S5 shows a typical behavior of their conductance as a function of *T*. Consistent with the small activation energy for proton transport in Nafion (<0.02 eV) [S11], we found that temperature effects in all the above devices are small over the entire temperature range (see Fig. S5). The nonmonotonic *T* dependence for the devices with Pt layers (Fig. S5) remains to be understood but we note that Nafion often exhibits similar nonmonotonic behavior at higher *T*, beyond the range of Fig. S5 [S14]. We speculate that the Pt activation shifts this peak to lower *T*. Importantly for our experiments, the influence of Pt nanoparticles on local conductivity in the Nafion constriction is approximately the same independently of whether an hBN membrane is present or not. This further indicates that the proton conductivity of Pt-activated hBN is so high that it becomes unmeasurable in our experimental setup, essentially because of the limited size of currently available hBN crystals.

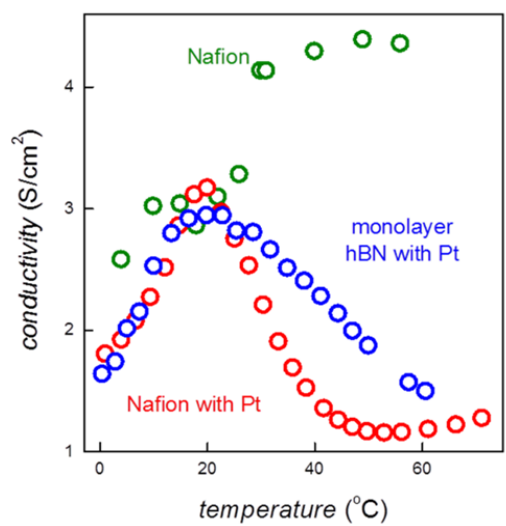

*Figure S5 | Proton transport limited by Nafion.* *Temperature dependences for bare-hole devices, Nafion/Pt/Nafion devices and membrane devices with catalytically-activated monolayer hBN. The nominal conductivity is calculated as the measured conductance S divided by the aperture area A.*

#3 Absence of atomic scale defects in 2D membranes

As described in section 1, visual inspection of membranes in SEM can reliably rule out holes and cracks with sizes down to <10 nm (see Fig. S2b). None of such defects could be found in many membranes we visualized in SEM. Occasional cracks such as in Fig. S2b could only be observed if introduced deliberately or a profound mistake was made during handling procedures. However, SEM cannot resolve nm- or atomic- scale defects, and other techniques are necessary to rule out microscopic holes. As already mentioned in the main text, no such defects have ever been reported for pristine graphene obtained by micromechanical cleavage in numerous SEM and scanning tunneling microscopy studies over many years. To add to this argument for the case of our particular membranes, we have used Raman spectroscopy that is known to be extremely sensitive to atomic-scale defects in graphene. The intensity of the D peak provides a good estimate for a concentration of such defects, which could be not only vacancies or larger holes but also adatoms that do not lead to pinholes. Despite our dedicated efforts, we could not discern any D peak in our graphene membranes. This sets an upper limit on the atomic defect density as ~$10^8$ cm$^{-2}$ or one defect per µm$^2$ [S15]. Furthermore, such a low density of defects in graphene is in stark contrast with a high density (~$10^{13}$ cm$^{-2}$) of sulfur vacancies found in mechanically cleaved MoS$_2$ [S16]. Notwithstanding, no proton current could be detected through our MoS$_2$ membranes. If we assume each vacancy to provide a hole of ~1 Å in size, the expected ~$10^5$ vacancies present in our typical MoS$_2$ membranes would provide an effective opening of ~30 nm in diameter. Using the results of Figure S3, this is expected to lead to a conductance of ~3 nS, that is, >100 times larger than the limit set by our measurements on proton conductance through monolayer MoS$_2$. This shows that individual vacancies in fact provide much smaller proton conductivity than their classical diameter suggests.

To strengthen the above arguments further, we tried to rule out even individual vacancies from our proton conductive (graphene and hBN) membranes. The most sensitive technique known to detect pinholes is arguably measurements of gas leakage from small pressurized volumes [S17,S18]. To this end, a microcavity of typically ~1 μm$^3$ in size is etched in a Si/SiO$_2$ wafer, sealed with graphene or hBN and then pressurized. If the pressure inside the microcavity is higher than outside, the membrane bulges upwards; if it is lower, downwards. Changes in pressure can be monitored by measuring the height of the bulge as a function of time using atomic force microscopy (AFM). If there are no holes in the membrane, the gas leaks slowly through the oxide layer, and it typically takes many hours until the pressure inside and outside the microcavity equalize [S17]. However, the presence of even a single atomic-scale hole through which atoms can effuse allows the pressure to equalize in less than a second [S18].

Following the procedures reported previously [S17,S18], we prepared microcavities in a Si/SiO$_2$ wafer and sealed them with monolayer graphene. The microcavities were placed inside a chamber filled with Ar at 200 kPa for typically 4 days to gradually pressurize them. After taking the devices out, the membranes were found to bulge upwards. Figure S6 shows the deflation of such micro-balloons with time. In agreement with the previous report [S17], the Ar leak rates were found to be ~10$^3$ atoms per second. If an atomic-scale hole is introduced by, for example, ultra-violet chemical etching, the leak rate increases by many orders of magnitude, leading to practically instantaneous deflation [S18]. Furthermore, we found no difference in the deflation rates for membranes with and without evaporated Pt.

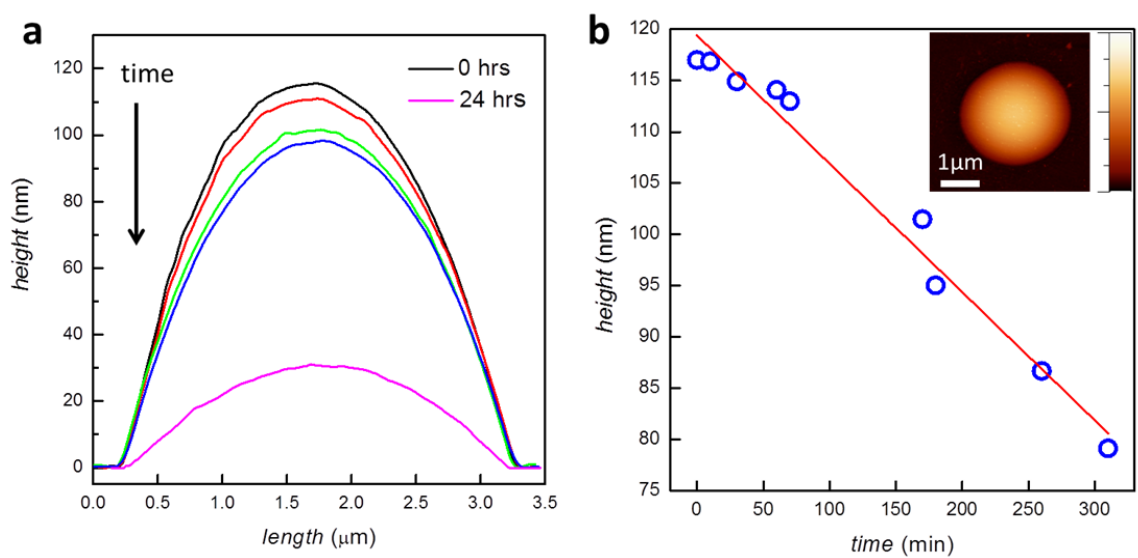

***Figure S6 | Deflation of micro-balloons to check for atomic scale defects in graphene membranes. a,** Height profiles for a typical graphene membrane at different times. **b,** Maximum height as a function of time. The inset shows a typical AFM image of a pressurized graphene microcavity (color scale: 0 to 130 nm). We measured six graphene membranes with all of them showing the same deflation rates, independently of whether Pt was deposited on top or not. Similar behavior was observed for hBN monolayers.*

In principle, it could be argued that membranes with pinholes smaller than the kinetic diameter of Ar (0.34 nm) or pinholes blocked with Pt nanoparticles should show no detectable leaks. However, monolayer membranes with subnanometer-sized pinholes are known to be rather unstable mechanically due to a tendency of defects to enlarge under strain [S18], which for the applied pressures reached significant values of ~1%. Our micro-balloons remained stable and could be pressurized many times. This behavior consistent with the previous work [S17,S18] assures that no individual pinholes were present in graphene and monolayer hBN obtained by mechanical cleavage.

#4 Detection of proton flow by mass spectrometry

To show directly that the electric current through our 2D membranes is carried by protons, we used an alternative setup described in the main text and shown in more detail in Fig. S7a. Protons transferring through graphene are collected at a catalyst Pt layer where they recombine to form molecular hydrogen: $2H^+ + 2e^- \rightarrow H_2$. The hydrogen flux is then measured with a mass spectrometer. Because the electric current *I* is defined by the number of protons passing through the graphene membrane, the hydrogen flow *F* is directly related to the passing current *I*, with no fitting parameters (see the main text).

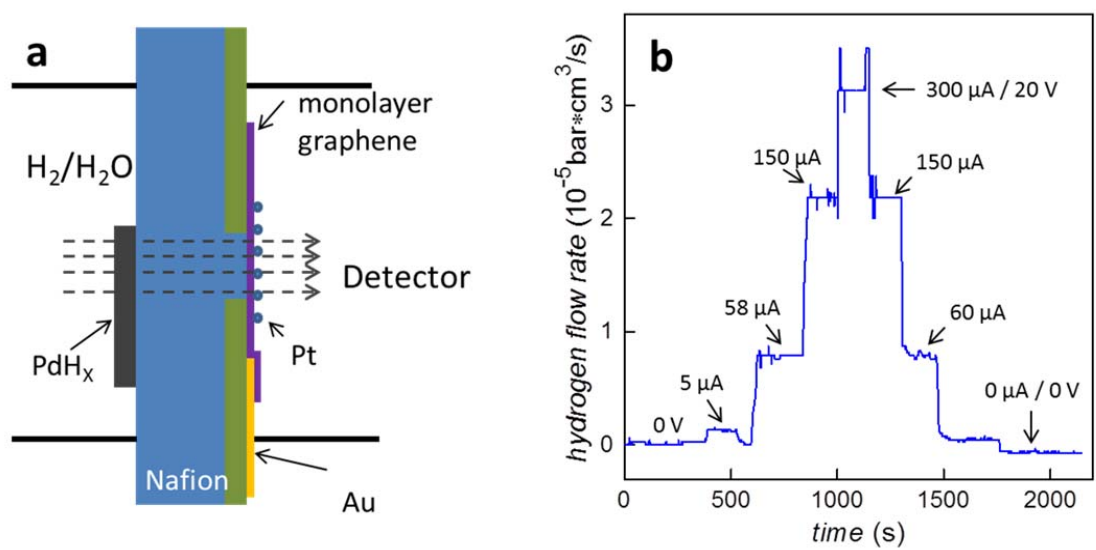

*Figure S7 | Hydrogen flow detection. **a**, Schematics of our devices for mass spectroscopy measurements. **b**, Example of the observed hydrogen flow rates as a function of time and measured current for different negative biases on the graphene membrane, which we applied in steps.*

For this particular experiment, the membrane devices were made as large as possible (50 μm in diameter), which was required to increase the hydrogen flux to such values that they could be detectable with our mass spectrometer (Inficon UL200). To collect the electric current at the graphene membrane, a metallic contact (100 nm Au/ 5nm Cr) was fabricated next to the $SiN_x$ aperture, before transferring graphene on top to cover both aperture and contact (right inset of Fig. 3 in the main text). This side of the Si wafer (with graphene on top) was then decorated with 1-2 nm of Pt to increase the proton flux and allow its easier conversion into hydrogen. The opposite face of the graphene membrane was covered with Nafion and connected to a $PdH_x$ electrode in the same way as described in section 1.

The resulting device on the Si wafer was glued with epoxy to a perforated Cu foil that was clamped between two O-rings to separate two chambers: one filled with a gas and the other connected to the mass spectrometer (Fig. S7a). First, we always checked the setup by filling the gas chamber with helium at the atmospheric pressure. No He leak could be detectable above background readings of the spectrometer at ~$10^{-8}$ bar cm$^3$/s. Then, the chamber was filled with our standard gas mixture (10% $H_2$ in argon at 1 bar and at 100% humidity). No hydrogen flux could be detected without applying negative bias to graphene. However, by applying such a bias a controllable flow of $H_2$ at a level of ~$10^{-5}$ bar cm$^3$/s was readily detected (Fig. S7b). This figure shows the hydrogen flow rates $F$ as a function of time for one of our devices using negative biases from 0 to 20 V. When cycling back from 20 to 0 V, the curves retraced themselves, indicating that the membrane was undamaged during the measurements.

Atomic hydrogen is highly unstable with respect to its molecular form, and it is most likely that the conversion into molecular hydrogen takes places at the surface of Pt rather than in the vacuum chamber. Accordingly, the Pt layer has to be discontinuous to let hydrogen escape. For continuous coverage (>5 nm of Pt), we observed formation of small hydrogen bubbles that grew with increasing the amount of electric charge passed through the circuit. Largest bubbles eventually erupted. It is also instructive to mention the case of continuous Au films evaporated on top of the above devices (already containing a discontinuous Pt layer). We found that a bias applied across such devices again resulted in the formation of bubbles at the interface between graphene and the metal film. The bubbles could burst and sometimes even damaged the membrane. This disallowed the use of continuous metal films for the mass spectrometry experiment. The same bubbling effect was observed for hBN membranes covered with a Pt film that provided the continuity of the electrical circuit for insulating hBN. These observations serve as yet another indication of proton transfer through graphene and hBN membranes. On the other hand, no bubbles could be observed for thicker 2D crystals that again shows their impermeability to protons. Note that, although thin Pd films were more transparent for atomic hydrogen (see the main text), large currents and hydrogen fluxes needed for mass spectrometry also led to hydrogen bubbles, which prevented the use of hBN with Pd in the mass spectrometry experiments.

**#5 Theoretical analysis of proton transport through 2D crystals**

As discussed in the main text, it is possible to understand our results qualitatively by considering the electron clouds created by different 2D crystals. These clouds impede the passage of protons through 2D membranes. In addition to the plots of the electron density for graphene and hBN monolayers in Fig. 1b of the main text, Figure S8 shows similar plots of these clouds with superimposed positions of C, B and N atoms using the ball-and-stick model of graphene and hBN crystal lattices. In addition, Figure S8 plots the electron density for monolayer $MoS_2$. One can immediately see that the latter cloud is much denser than those of monolayer hBN and graphene, which explains the absence of proton transport through $MoS_2$ monolayers.

For quantitative analysis, we first note that proton permeation through graphene has previously been studied using both ab initio molecular dynamics simulations (AIMD) and the climbing image nudged elastic band method (CI-NEB) [S19-S21]. These studies have provided estimates for the proton transport barrier $E$ in graphene ranging from ≈1.17 eV to 2.21 eV [S19-S21]. We reproduced those results for the case of graphene and extended them onto monolayer hBN.

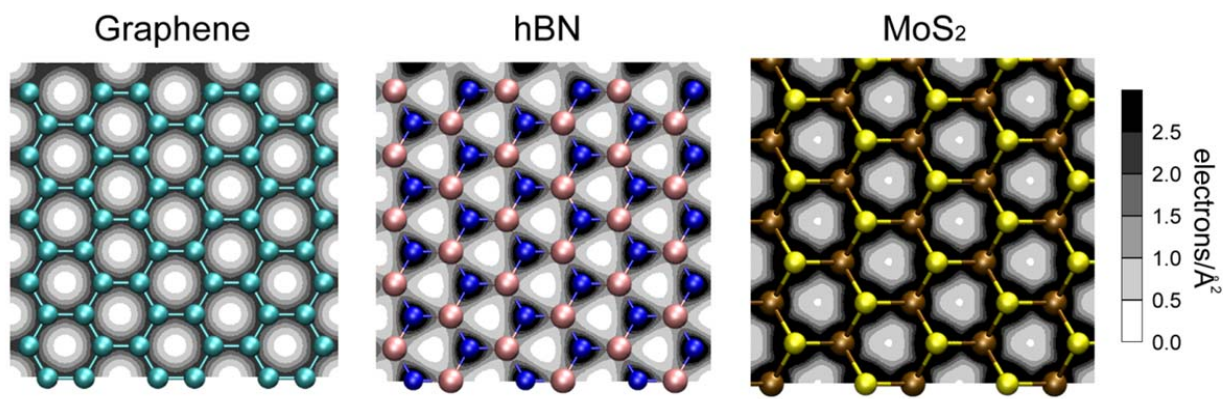

*Figure S8 | Electron clouds of 2D crystals.* *Integrated charge densities for graphene, monolayer hBN (nitrogen is indicated by blue balls; boron in pink) and monolayer MoS$_2$ (S is in yellow, Mo in brown).*

All our simulations were performed using the CP2K package [S22] with the Pade exchange-correlation functional form [S23]. The energy cutoff of plane-wave expansions was 380 Ry, and we used the double-$\zeta$ valence basis with one set of polarization functions [S24] and the Goedecker-Teter-Hutter pseudopotentials [S23]. In the first approach, the bombardment of graphene and monolayer hBN with protons of varying kinetic energy was simulated using AIMD in the NVE ensemble (that is, Number of atoms, Volume and Energy are assumed constant). The barrier was estimated as the minimum kinetic energy necessary for proton transfer. The AIMD simulations have yielded graphene's *E* between 1.30 eV and 1.40 eV, in agreement with results of refs S19,S20.

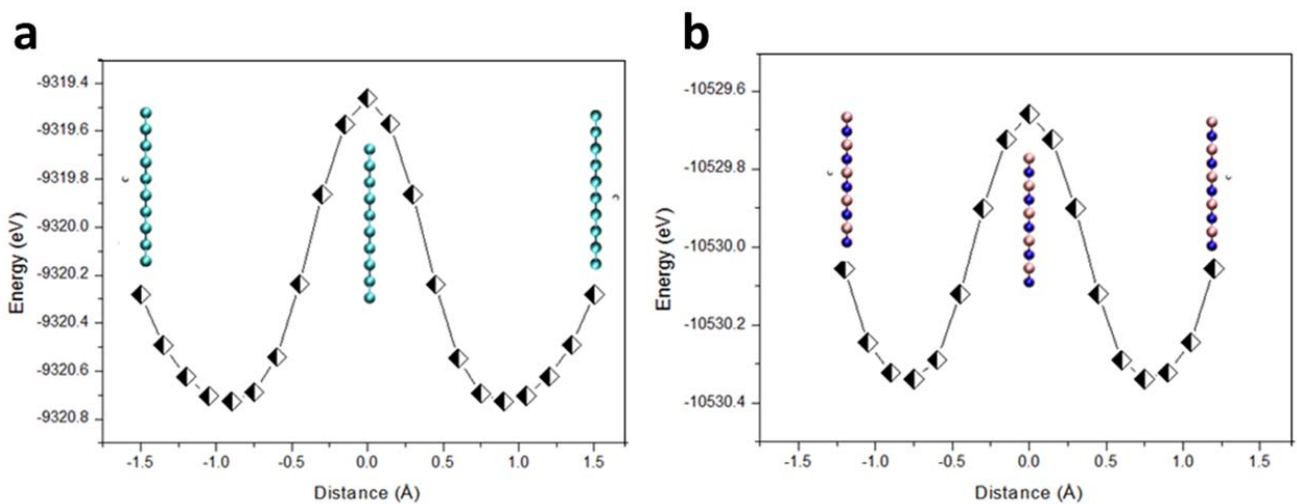

*Figure S9 | CI-NEB simulations.* *Energy profiles as a function of the proton distance to the center of the hexagonal ring in graphene and hBN (**a** and **b**, respectively). Carbon atoms are shown as cyan-colored spheres, nitrogen in blue, boron in pink and protons (H$^+$) in white.*

In the second (CI-NEB) approach, we calculated the energy for various configurations (usually referred to as 'images'), which correspond to different distances between a proton and a 2D membrane [S25]. This provided a series of images for a proton approaching the membrane. The energy was then minimized over obtained images and plotted as a function of distance to 2D crystals. The barrier *E* was estimated using the differential height of energy profiles. Figure S9 shows examples of such energy profiles for graphene and monolayer hBN. From the CI-NEB calculations, we have estimated the proton barrier as 1.26 eV and 0.68 eV for graphene and monolayer hBN, respectively, in agreement with the AIMD results.Finally, to model the effect of Pt on proton transport, we again used AIMD simulations. To this end, 4 Pt atoms were placed at a fixed distance (4 Å) from the graphene membrane and the bombardment with protons was simulated as described above. The addition of the Pt atoms resulted in a significant reduction of the barrier to ≈0.6 eV; that is, by a factor of 2. The absolute value of the reduction in the barrier height is in good agreement with the experimental observations.

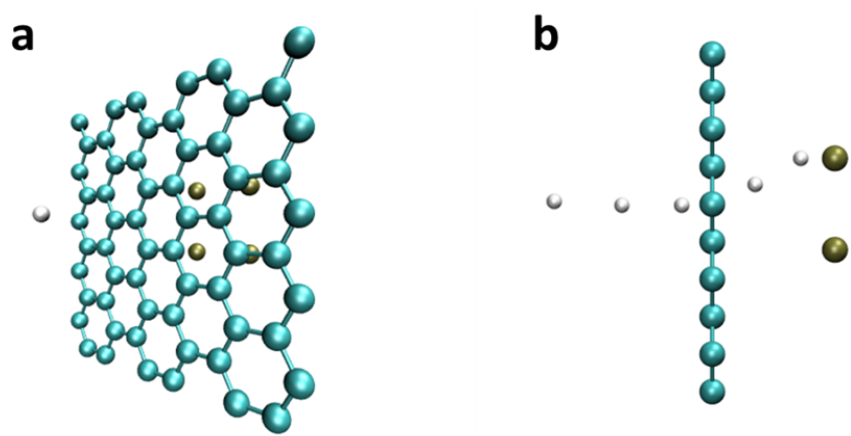

***Figure S10 | AIMD simulations for the proton barrier in graphene with Pt.*** *Carbon atoms are shown in cyan, Pt in ochre, $H^+$ in white.* ***a****, Experimental situation is mimicked by placing 4 Pt atoms at a distance of 4 Å from the graphene sheet.* ***b****, Figure shows the trajectory of protons with initial kinetic energy E =0.7 eV (the other two Pt atoms cannot be seen due to the perspective). The bended trajectories indicate that the decreased barrier is due to the interaction of protons with Pt.*

#6 Proton transport through 2D crystals in liquids

Although Nafion was the material of choice in this work due to its stability and convenience of handling, in order to show the generality of our results, we have also investigated proton conductivity of 2D crystals when they were immersed in water solutions. For these experiments, devices were fabricated in the same way as described previously but instead of covering 2D crystals with Nafion, they separated two reservoirs containing liquid electrolytes (HCl solutions). A polydimethylsiloxane seal was used to minimize leakage along the 2D crystal/substrate interface (Fig. S11 inset; yellow) [S9]. Ag/AgCl electrodes were placed in each reservoir to apply a bias across the membranes and measure ionic currents (Fig. S11).

Typical I-V profiles of single-, bi-, and tri- layers hBN are presented in Fig. S11a. This behavior was highly reproducible as evidenced by the statistics in Fig. S11b. For devices prepared in the same manner but without a 2D crystal, the conductivity *S* was >$10^4$ times higher than in the presence of monolayer hBN, which ensured that the 2D crystals limited the proton current. As in the case of Nafion, we found a parasitic parallel conductance but it was somewhat higher (~20 pS) for the liquid cell setup. Within this accuracy, we could not detect any proton current through monolayer $MoS_2$, bilayer graphene, trilayer hBN or any thicker 2D crystals. Most importantly, the measured proton conductivities using electrolytes agree extremely well with the values found using Nafion as the proton conducting medium (cf. Fig. 1b and Fig. S11b).

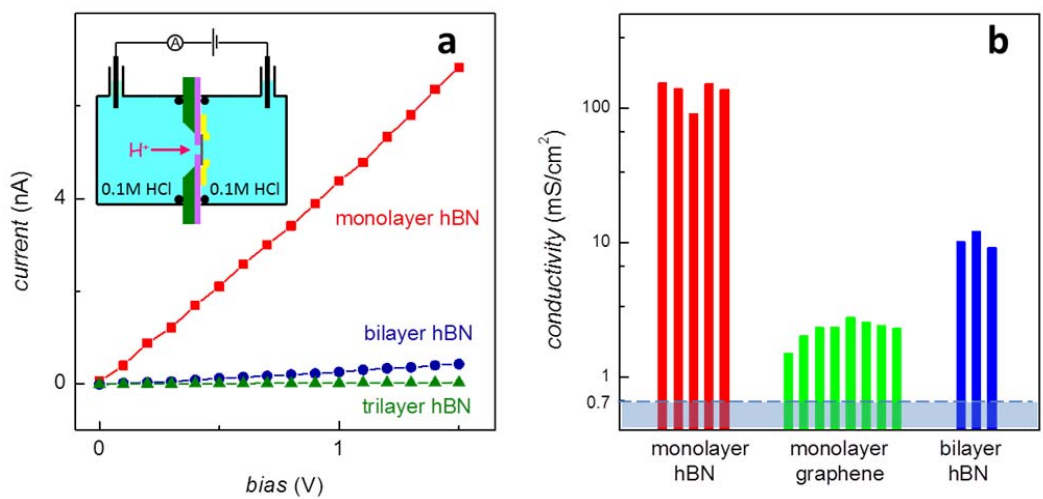

*Figure S11 | Proton transport through 2D crystals in liquids. **a**, Examples of I-V characteristics for mono-, bi- and tri-layer hBN covering an aperture of 2 μm in diameter. The inset shows schematics of the liquid cell. In the case of trilayer hBN, the current is within the range given by a parasitic parallel resistance. **b**, Histograms for the 2D crystals that exhibited clear proton current in the liquid cell setup. Each bar represents a different sample with a 2 μm diameter membrane. The shaded area shows the detection limit set by leakage currents.*

**Supplementary references**

[S1] K. S. Novoselov *et al*. Two dimensional atomic crystals. *Proc. Natl Acad. Sci. USA* **102**, 10451-10453 (2005).
[S2] R. R. Nair *et al.* Graphene as a transparent conductive support for studying biological molecules by transmission electron microscopy. *Appl. Phys. Lett.* **97,** 153102 (2010).
[S3] A. V. Kretinin *et al.* Electronic properties of graphene encapsulated with different two-dimensional atomic crystals. *Nano Lett.* **14,** 3270–3276 (2014).
[S4] D. W. Murphy *et al.* A new route to metal hydrides. *Chem. Mater.* **5,** 767–769 (1993).
[S5] K. Mauritz, R. B. Moore. State of understanding of nafion. *Chem. Rev.* **104,** 4535–4585 (2004).
[S6] H. Morgan, R. Pethig, G. T. Stevens. A proton-injecting technique for the measurement of hydration-dependent protonic conductivity. *J. Phys. E* **19**, 80–82 (1986).
[S7] C. Zhong *et al.* A polysaccharide bioprotonic field-effect transistor. *Nat. Commun.* **2,** 476 (2011).
[S8] S. Schuldiner, G. W., Castellan, J. P. Hoare. Electrochemical behavior of the palladium-hydrogen system. I. Potential-determining mechanisms. *J. Chem. Phys.* **28,** 16-19 (1958).
[S9] S. Garaj *et al.* Graphene as a subnanometre trans-electrode membrane. *Nature* **467,** 190–1933 (2010).
[S10] N. Agrait, A. L. Yeyati, J. M. van Ruitenbeek. Quantum properties of atomic-sized conductors. *Phys. Rep.* **377**, 81-279 (2003).
[S11] Y. Sone, P. Ekdunge, D. Simonsson. Proton conductivity of Nafion 117 as measured by a four electrode AC impedance method. *J. Electrochem. Soc.* **143,** 1254–1259 (1996).